\documentclass{emulateapj} 
\newcommand{\myemail}{kregenow@ssl.berkeley.edu}

\shorttitle{FUV Obs of O-E Superbubble}
\shortauthors{Kregenow et al.}

\begin{document}

\title{Far-UV Observations of a Thermal Interface in the Orion-Eridanus Superbubble}

\author{
J. Kregenow\altaffilmark{1},
J. Edelstein\altaffilmark, E. Korpela,
B. Welsh, C. Heiles\altaffilmark{2},
K. Ryu\altaffilmark{3}, K. Min,
Y. Lim, I. Yuk\altaffilmark{4},
H. Jin, K. Seon
}

\slugcomment{{\it Accepted for Publication in ApJL.} { Contact: }{\myemail}}

\altaffiltext{1}{Space Sciences Lab, Univ of CA, Berkeley, CA 94720} 
\altaffiltext{2}{Astro Dept, Univ of CA, Berkeley, CA 94720}  
\altaffiltext{3}{Korea Advanced Inst of Sci \& Tech, 305-701, Daejeon}
\altaffiltext{4}{Korea Astro \& Space Sci Inst, 305-348, Daejeon, Korea}


\begin{abstract}
Diffuse far-UV emission arising from the edge of the
Orion-Eridanus superbubble was observed
with the \emph{SPEAR} imaging spectrometer, revealing
numerous emission lines arising from both atomic species
and H$_{2}$. Spatial variations in line intensities of
C{\sc iv}, Si{\sc ii}, and O{\sc vi}, in comparison with
soft X-ray, H${\alpha}$ and dust data, indicate that
these ions are associated with processes
at the interface between hot gas inside
the bubble and the cooler ambient medium.
Thus our observations probe physical conditions
of an evolved thermal interface in the ISM.

\end{abstract}


\section{Introduction}

\begin{figure}
\epsscale{.4}
\plotone{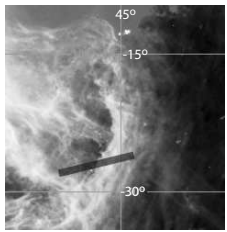}
\caption{Composite 1/4~keV, H$\alpha$, and dust map image
\citep{skyview} of an OES interface
in greyscale [color image online], overlaid with \emph{SPEAR}
slit position. The \emph{SPEAR} field spans
the edge of the OES wall, sampling both hot inside and cooler  
outside material. The gridlines show ecliptic latitude and longitude.}
\label{slitpositions}
\end{figure}

The Orion-Eridanus Superbubble (OES) is a large cavity
in the interstellar medium (ISM)
created by some combination of
stellar winds and supernovae from the enclosed
Orion OB1 stellar association, \citep[e.g.][]{reynolds79}.
Its comparable proximity and size
($\sim$300~pc) give it a large apparent angular
extent of $\sim$30$^{\circ}$ \citep{burrows96},
allowing for detailed study of its structure. 
The complicated morphology and dynamics, described by \cite{guo95}, 
indicate that the OES is still evolving and expanding into the ISM
and thus it likely includes a variety of interfaces and
asymmetries resulting from interacting phases of interstellar gas.
The OES has clearly-delineated thermal boundaries 
between the ambient ISM and the suberbubble cavity.
The cavity is bright in soft X-ray (SXR) emission, indicating 
that it is filled with hot T$\sim$$10^{6}$K gas
\citep[see][]{heiles99}.
For example, the X-ray edge of the OES toward Galactic South
(lower right of Fig. \ref{slitpositions})
coincides with a ridge of H$\alpha$
emission produced by cooler T$\sim$$10^4$~K gas, 
beyond which still-cooler neutral H and dust 
are seen in 21-cm \citep{heiles99} and infrared \citep{burrows93}.

We took a detailed far ultraviolet (FUV) emission spectrum with the 
\emph{SPEAR} instrument
to study the physical conditions across one edge of the OES,
where hot and cooler gas  are likely to be interacting.
{\em SPEAR} (aka {\em FIMS}) spectrally images
diffuse FUV background radiation with
\ensuremath{\lambda /\Delta \lambda}$\sim$550, using
two large field of view (FOV) imaging spectrographs
optimized for measuring diffuse emission:
the Short 'S' (900-1150 {\AA})
and Long 'L' (1350-1750 {\AA})
channels.
The instrument and on-orbit performance are described
in \citet{edelstein05a}.\\

\section{Observations and Data Reduction}

\emph{SPEAR} observed a fixed location in the OES
where hot X-ray emitting gas abuts cooler H$\alpha$ and
infrared (IR) emitting gas. The imaged field
(see Fig.\ref{slitpositions})
spans the bright outermost
H$\alpha$ front called 'Arc~B' by \citet{boumis01},
and is large enough to simultaneously measure 
both the hot X-ray region inside
and the colder IR region outside the bubble.
The target was observed over 60 orbits in Dec 2003
for a cumulative exposure time of 31~ks.

The data were processed as described in \cite{edelstein05b}.
We removed data recorded at times with high count rate
($>$$10\mbox{ s}^{\mbox{-}1}$ for S and $>$$20\mbox{ s}^{\mbox{-}1}$
for L channel) to mitigate
terrestrial airglow contamination.
Spectra were binned by 1~\AA, smoothed by a 3~\AA \ running boxcar,
and fluxed to Line Units (LU, photons s$^{-1}$cm$^{-2}$sr$^{-1}$).
All of the L channel photons were mapped to the sky and
the image inspected for the presence of stars.
Two $<$20$\arcmin$-wide image features are consistent with
stellar profiles, with a flux of
$\sim$$3 \mbox{-} 5$$\times$$10^{-14}$~erg s$^{-1}$cm$^{-2}$\AA$^{-1}$.
If the features are indeed
caused by faint UV stars, then they contribute no more than
7\% of the background continuum flux over 1$^{\circ}$ --
an unimportant source of error in our spectral line analysis.
Potential stellar contributions to the S channel are even less significant.\\

\section{Spectral Analysis}
\label{spectra}

The raw spectra show a strong underlying
background component in addition
to the many emission lines of interest. 
We used a spectral model \citep{korpela05model}
to simultaneously fit the
background and overlying emission lines. 
The background model has an equivalent total intensity of 3500
continuum units (CU, photons s$^{-1}$cm$^{-2}$sr$^{-1}$\AA$^{-1}$) in the
S channel and 1000~CU in L.
The background signal primarily consists of contributions from
instrumental dark noise (80$\%$ of the background in S channel) and
starlight scattered by interstellar dust ($\sim$60$\%$ in L),
with instrumentally-scattered geocoronal Ly-$\alpha$ emission also
contributing to the S background.
The starlight is fit by a canonical power-law (index=0.26)
stellar luminosity function 
which is absorbed and scattered by intervening dust and gas
with N(H{\sc i})$\simeq$$4$$\times$$10^{20}$cm$^{-2}$, 
a value similar to the
N(H{\sc i})$\simeq$$5$$\times$$10^{20}$cm$^{-2}$
measured in this location by the
Leiden/Dwingeloo Survey \citep{hartmann97}.
We subtract the fitted background model
from the raw spectrum to obtain a net emission spectrum.
Many emission lines are apparent in this net spectrum.
The strongest are terrestrial in origin -
arising from airglow in Earth's upper atmosphere -
including the bright Lyman series from 912-1026 \AA, O{\sc i}
at 990/1041 \AA, some fraction of N{\sc i} at 1133 \AA,
and O{\sc i} prominently at 1356 \AA\
and possibly blending with He{\sc ii} at 1641~\AA.
We identify and attribute most of the remaining lines to
astrophysical sources, namely originating in the OES or ISM.

\begin{figure}
\includegraphics[scale=.45,angle=90]{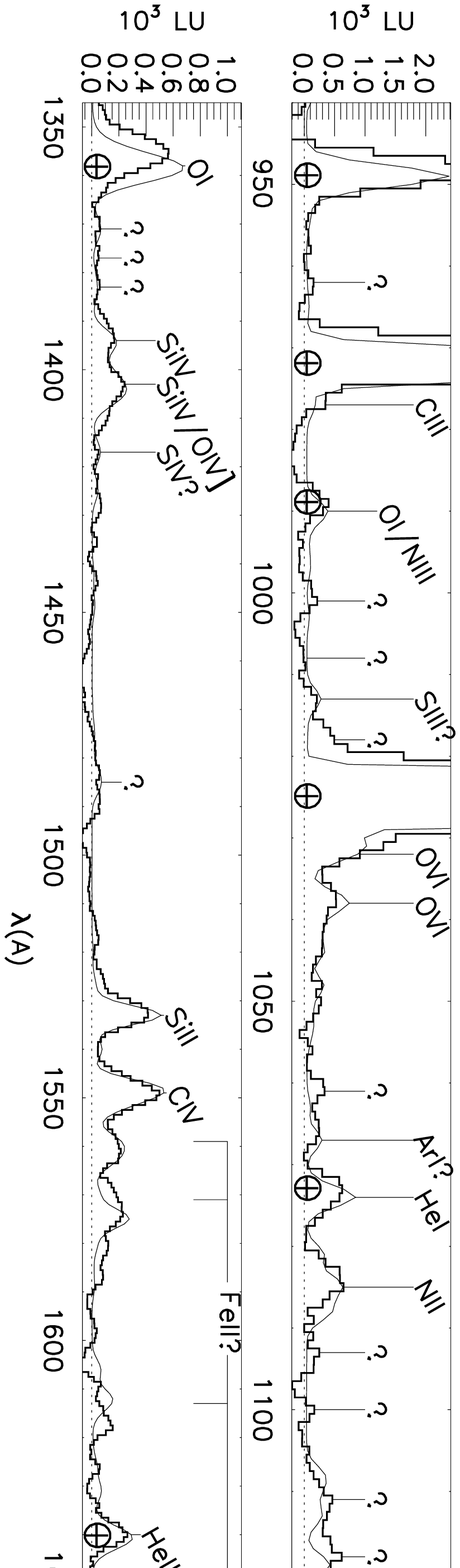} 
\caption{Fluxed \emph{SPEAR} spectra (thick line)
after subtraction of modeled continuaa and H$_{2}$ are
overplotted with a modeled composite spectrum (thin line) of both
astrophysical and atmospheric airglow emission lines. Geocoronal airglow
lines are indicated by the Earth symbols, including 4th order He{\sc i} 537 \AA.}
\label{SLspectra}
\end{figure}

Several significant emission features could not be accounted for
by neutral or ionized atomic species. 
A theoretical spectrum of H$_{2}$ fluorescence lines 
(Draine 2004, private  communication),
however, provides a good fit to most of these lines and blends. 
The total H$_{2}$ flux is 19~kLU
in S channel, and 32~kLU in L. This is $\sim$10$\%$ of
the integrated flux from the scattered
stellar continuum fit, whose band-averaged value is
$\sim$400 and $\sim$570 CU in the S and L channel,
respectively.
Thus we have discovered a significant H$_{2}$ component
toward the OES. We defer discussion of this 
component to future work; see \citet{ryu05} for an analysis of the
physical parameters and proposed location and origin of the H$_{2}$
gas detected by \emph{SPEAR} over the entire Eridanus region.
For the present analysis, we subtracted the modeled H$_{2}$ component
to obtain a net atomic emission spectrum, shown in Fig. \ref{SLspectra}.


We find lines from species with a wide range of ionization potentials
(6-114~eV),
indicative of both cool and hot interstellar gas along the sight-line. 
The spectral model \citep{korpela05model} was used to estimate the
emission line strengths. 
Some of the most significant (S/N$>$3) astrophysical line model
detections 
are O{\sc vi} and N{\sc ii} in S channel, 
and Si{\sc iv}, O{\sc iv]}, Si{\sc ii}, C{\sc iv}, and Al{\sc ii} in L channel.
A list of identified atomic lines, observed line centers to the nearest
1~\AA, total line flux, and signal to noise ratio
is shown in Table \ref{linetable}. 
The spectral model finds traces of other lines including
Fe{\sc ii}, Ar{\sc i} and S{\sc iii}, along with other undetected lines.
Before these identifications can be made secure, however,
more careful analysis is required to account for potential confusion 
from H$_{2}$ fluorescence and airglow lines.
Such analysis was done for C{\sc iii} (977\AA), which is close to the
Ly$\gamma$ (973\AA) airglow line. We only establish
an upper limit (90\% confidence) to the C{\sc iii} flux, which is included
in the Table \ref{linetable}.

\begin{deluxetable}{lccc}
\tablewidth{0pt}
\tabletypesize{\scriptsize} 
\tablecolumns{4}
\tablecaption{Modeled Emission Lines in the OES}
\tablehead{\colhead{Wavelength} &  
\colhead{Species} &
\colhead{Intensity} & \colhead{S/N } \\
\colhead{$\lambda$(\AA)} & \colhead{ID} &
\colhead{($10^3$LU)} & \colhead{}}
\startdata
   977      & C{\sc iii}  &    $<$3.6  &    1.7    \\ 
   990\tablenotemark{a,b} & O{\sc i} &     0.2  &    0.4    \\
   990\tablenotemark{b} & N{\sc iii} &     0.9  &    1.6    \\
  1013       & S{\sc iii}? &     0.7  &    1.3    \\
  1032        & O{\sc vi} &     2.0  &    3.4    \\
  1038        & O{\sc vi} &     1.0  &    1.8    \\
  1067       & Ar{\sc i}? &     0.8  &    1.5    \\
  1074\tablenotemark{a}        & He{\sc i} &     2.4  &    4.2    \\
  1085        & N{\sc ii} &     1.8  &    3.0    \\
  1134\tablenotemark{a}        & N{\sc i} &     5.3  &    7.2    \\

  1358\tablenotemark{a}         & O{\sc i} &     3.6  &   17.1    \\
  1394       & Si{\sc iv} &     0.9  &    5.5    \\
  1403\tablenotemark{b}       & Si{\sc iv} &     0.4  &    2.8    \\
  1404\tablenotemark{b}       & O{\sc iv]} &     0.7  &    4.2    \\
  1417       & S{\sc iv}? &     0.2  &    1.6    \\
  1533       & Si{\sc ii} &     2.1  &   13.0    \\
  1549        & C{\sc iv} &     2.2  &   12.7    \\
  1640\tablenotemark{b}       & He{\sc ii} &     0.5  &    2.0    \\
  1641\tablenotemark{a,b} & O{\sc i} &     0.9  &    3.4    \\
  1657         & C{\sc i} &     0.6  &    2.3    \\
  1671       & Al{\sc ii} &     1.9  &    6.3    \\
\enddata
\tablenotetext{a}{Attributed to Airglow}
\tablenotetext{b}{Blended Line}
\label{linetable}
\end{deluxetable}

\subsection{Spatial Variation of FUV Emission Intensity}
\label{spatialdiscussion}
In order to understand the origin of the observed FUV emission-line species,
we have examined the spatial variation of emission
along the long dimension of the \emph{SPEAR} FOV
for three bright emission lines that trace species of very different
ionization states,
and compared the spatial variations to other bands.
The emission lines used in this analysis are
O{\sc vi} (1032~\AA), C{\sc iv} (1549~\AA), and Si{\sc ii} (1533~\AA).
Energies of 114, 48, and 8~eV are required to produce
O{\sc vi}, C{\sc iv}, and Si{\sc ii}, respectively.

For the C{\sc iv} and Si{\sc ii} analysis, we divided the L-channel
FOV into 1.0$^{\circ}$ bins and made separate spectra for each location.
The C{\sc iv} and Si{\sc ii} intensity were clearly diminished
in some locations,
so we defined a baseline spectrum with relative intensity zero in
the location where both lines were weakest. We then subtracted this
baseline C{\sc iv}/Si{\sc ii} spectrum from the other spectra
after scaling to fit
the continuum in a quiet region (around 1500 \AA).
Assuming that the continuum
did not change between FOV locations, scaling and subtracting
the baseline spectrum leaves only the excess line emission at each location.
We then fit Gaussian line profiles to the residuals to measure the
C{\sc iv} and Si{\sc ii} excess intensity.
There is little if any Si{\sc ii} emission
in the baseline spectrum, so we assume that the excess intensity is
similar to the absolute intensity for Si{\sc ii}.
We note that the absolute intensity of C{\sc iv}, however,
may be significantly underestimated
because there is a strong, broad C{\sc iv} absorption line 
in the scattered stellar background
not taken into account here.

\begin{figure}
\epsscale{.9}
\plotone{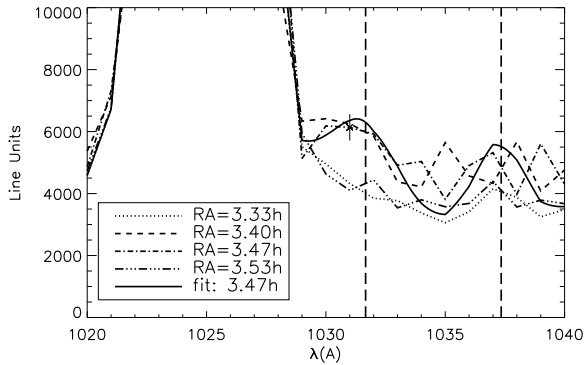}
\caption{\emph{SPEAR} spectra, zoomed in on the O{\sc vi} doublet
in the wings of Ly$\beta$,
for 4 locations spanning the OES edge
with a typical error bar at 1031 \AA.
The smooth solid line is a Gaussian fit to the spectrum at
RA=3.47h, where O{\sc vi} is clearly enhanced.
}
\label{OVIprofile}
\end{figure}

We measured O{\sc vi} relative intensity variations using a
similar approach.
The O{\sc vi} (1032~\AA) line is resolved, but close to the intense
Ly$\beta$ airglow line (1026~\AA), which dominates all features in
the S channel. So instead of scaling to the continuum, we performed a 
near-unity scaling to fit the short wavelength shoulder of the
Ly$\beta$ line profile. Fig \ref{OVIprofile} shows the spectra
for the four locations after this scaling but before subtraction.
Note how the O{\sc vi} intensity is clearly enhanced in two adjacent
locations. (At the distance of the OES, this 1$\sim$2$^{\circ}$ zone
corresponds to 5$\sim$10~pc.)
Assuming that the \emph{airglow line profile}
did not change between locations, scaling to Ly$\beta$ and subtracting
the baseline leaves only the excess O{\sc vi} emission at each location.
Again there is little if any O{\sc vi} emission in the minimum
location, so we assume that the excess intensity is approximately equal
to its absolute intensity.\\

\section{Discussion}

The global structure and coherence of the OES is a matter of some
debate. In particular, it has been suggested that one of the two prominent
H$\alpha$ filaments in the region ('Arc~A', see \cite{boumis01}) may be part
of a more distant shell structure unrelated to the OES.
But Arc~A only grazes the edge of \emph{SPEAR's} FOV where we measure no
enhancement, while Arc~B's FUV enhancement is quite
prominent in the center of the FOV. Since the features are unconfused
and the contribution from Arc~A is minimal,
we presume that our measurements are dominated by Arc~B.

The present \emph{SPEAR} observations
reveal interstellar gas with a wide range of
ionization toward the OES whose
origin and location we wish to clarify.
The emission line species
detected range from low ionization state ions (e.g. Si{\sc ii}, Al{\sc ii})
which can be created entirely by photo-ionization
of the cold neutral ISM by the ambient interstellar radiation field,
to species (e.g. Si{\sc iv}, C{\sc iv}) that can be created either by
photo-ionization or by thermal ionization,
to species (e.g. O{\sc vi}, He{\sc ii}, O{\sc iii]})
likely created only by thermal excitation because of their
high ionization potentials or low optical depth.

The OES has been previously observed
twice in the UV.
\cite{paresce83} observed regions of the OES near our target with
a narrow-band photometer and concluded that their measured increase
in the UV count rate over the bubble edge was most likely
due to some combination of H$_{2}$ fluorescence,
thermal, or H{\sc i} two-photon emission from ionized gas.
Two small, $<$1$^{\circ}$, fields (within
5$^{\circ}$ of our target but not spanning the bubble edge) were observed
with the $\it Voyager$ low-resolution spectrometer \citep{murthy93}.
They attributed a strong FUV
continuum to starlight scattered by interstellar dust
with possible small additional contributions from unidentified emission
lines and two-photon continuum emission from H{\sc i} radiative decay.
These previous data were insufficient to detect
spectral line variation across the bubble interface, leaving
questions unanswered about
how the FUV emission is linked to the physical properties of the
bubble's interface with the ambient interstellar media.

\begin{figure}
\epsscale{0.9}
\plotone{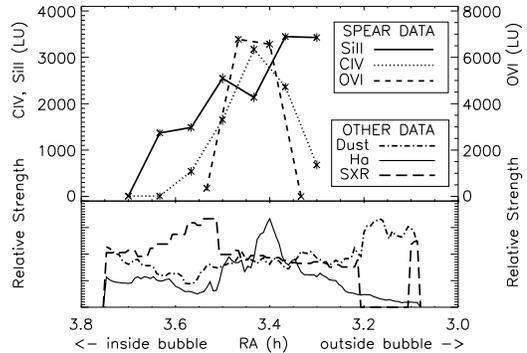}
\caption{
Spatial Variation of emission
along \emph{SPEAR}'s FOV.
Emission line intensity in Line Units vs. sky position is plotted
for O{\sc vi} (1032~\AA), C{\sc iv}, and Si{\sc ii} emission in 1$^{\circ}$ bins.
For comparison, the lower panel shows relative (scaled)
Dust, H$\alpha$, and 1/4~keV x-ray emission. The sharp peak in H$\alpha$
at RA=3.4h is the expanding bubble edge.}
\label{spatialvariation}
\end{figure}

The spatial variation analysis results are shown in the upper panel
of Fig. \ref{spatialvariation},
which plots line strength of the three analyzed species
per one-degree bin along the FOV:
four bins in S channel, and seven in L.
The high-ionization lines O{\sc vi} and C{\sc iv} are
expected to trace high-temperature gas, while the low-ionization
Si{\sc ii} line should trace lower-temperature gas and dust.
Since the
\emph{SPEAR} FOV spans a thermal boundary in the OES,
separate spectra sample locations both inside
and outside the superbubble.
The lower panel shows spatial variation in emission from three other
wavebands for comparison: 1.4~keV SXR \citep{rosatASS},
dust \citep{schlegel98}, and H$\alpha$ \citep{finkbeiner03}.  (Data
obtained from NASA's SkyView facility, http://skyview.gsfc.nasa.gov).
We note that both C{\sc iv} and O{\sc vi} intensity are highest at
RA= 3.45h, where H$\alpha$ emission peaks.
This behavior strongly supports the inference 
that the majority of C{\sc iv} and O{\sc vi} emission originates
from the OES thermal interface.
The ambient media outside the bubble
are likely far too cold to produce O{\sc vi}, while the 
hot, thin X-ray emitting gas inside has 
too low of an O{\sc vi} or C{\sc iv} ionization fraction and
emission measure to produce measurable emission.
The Si{\sc ii} emission,
in contrast, increases \emph{outside} the bubble 
where the enhanced dust reddening map 
indicates cool ambient gas resides.

Whether the interface contains a fast interstellar shock
is an important question about its nature.
\cite{hartigan87} provide a means
for probing this using the O{\sc vi}/C{\sc iii} (977\AA) flux
ratio, a sensitive indicator of shock velocity.
In an observation a few degrees away from ours at a similar location
on the H$\alpha$ interface, \cite{murthy93}
detect C{\sc iii} at 20~kLU but do not robustly detect O{\sc vi},
at 10$\pm$10~kLU. Our firm 3~kLU detection of O{\sc vi} is consistent with
their result, being well below their detection threshhold, but
our C{\sc iii} upper limit is much smaller at 3.6~kLU.
The O{\sc vi}/C{\sc iii} ratios from \emph{Voyager} and \emph{SPEAR} data,
$<$1 and $>$0.8 respectively, could arise from a shock with velocity
140-180~km/s according to the Hartigan models.
However, \citet{naranan76} compared the OES to Vela and Cygnus  -- two
middle-aged supernova remnants known to contain fast shocks --
and found Eridanus to be older, larger, and
fainter in X-ray emission, concluding that if the OES has a
supernova origin then it is more evolved with much lower velocities.
Additionally, the SPEAR observations of
Vela \citep{nishikida05} and Cygnus \citep{seon05}
show two orders of magnitude more OVI and CIII emission
than Eridanus, and have OVI/CIII ratios between one and two.
Moreover, there are
no direct observations of high-velocity gas in the Eridanus region
that could produce highly-ionized species such as O{\sc vi} by
thermal heating from fast shocks.
Both \citet{heiles99} and \citet{reynolds79}
showed the OES H{\sc i}/H$\alpha$ shell
to be expanding at $<$20~km/s, and
interstellar absorption observations \citep{welsh05}
show no high-velocity absorption components.
Finally, if the \emph{Voyager} and \emph{SPEAR}
measurements are sampling a similar interface,
then it is puzzling that the \emph{Voyager}
measurement of C{\sc iii} so far exceeds the \emph{SPEAR} upper limit.
We believe the difficulty in assesment of background subtraction and
separating Ly$\gamma$ from C{\sc iii} in \emph{Voyager} data could have
compromised the result and explain the discrepancy.

We conclude that a fast shock is one possibility at the interface, but
there is not a plurality of evidence to support it. We suggest that
a slow shock or a 'quiescent' (non-shock) thermal interface between
the hot X-ray producing interior cavity of the OES
and the cool ambient medium could be responsible for
producing the highly-ionized species we observe. \\

\section{Conclusions}
\label{conclusions}

We have discovered numerous diffuse atomic and molecular FUV emission lines
emanating from both hot and cool gas toward the OES.
Of these lines, only O{\sc vi} has been previously detected in the OES at all,
though not at the bubble edge.
We measured intensity variations of O{\sc vi}, C{\sc iv}, and Si{\sc ii}
emission across the interface between the hot bubble interior and the
ambient cooler media.
The O{\sc vi} and C{\sc iv} emission are greatly enhanced at the interface,
and clearly show an ionized stratification.
While these ions could be formed in a fast shock,
we suggest that a quiescent thermal interface model is also
consistent with the FUV observations and with other
previous observations finding only low-velocity gas.
We have also observed a significant H$_{2}$ component,
despite the moderate total hydrogen column in this direction.

The spatial variation of emission line intensity across the interface
holds much promise for probing the physics of the interface,
especially when applied to
species of widely varying ionization potential.
To further probe the physics of the OES region,
we intend to extend our analysis to additional spectral lines
and to two more deep observations \emph{SPEAR} made of the OES
toward nearby fields.
We will also
test interface models predicting the spectral components.
These new data can contribute to the understanding of interstellar 
thermal heating sources and interfaces by 
comparing the FUV emission line observations with 
predictions made with interface models invoking shocks,
conduction, and turbulent mixing.

\acknowledgments
\emph{SPEAR/FIMS}
is a joint project of KASSI, KAIST, and U.C. Berkeley,
funded by the Korea MOST and NASA Grant NAG5-5355.
J. Kregenow is supported by the National Physical Sciences  
Consortium Fellowship, and C. Heiles in part by NSF grant
AST 04-06987.

\bibliographystyle{apj}

\begin{thebibliography}{22}
\expandafter\ifx\csname natexlab\endcsname\relax\def\natexlab#1{#1}\fi

\bibitem[{{Boumis} {et~al.}(2001)}]{boumis01}
{Boumis}, P. {et~al.} 2001, \mnras, 320, 61

\bibitem[{{Burrows} \& {Guo}(1996)}]{burrows96}
{Burrows}, D. \& {Guo}, Z. 1996, in Roentgenstrahlung from the Universe,
  221--224

\bibitem[{{Burrows} {et~al.}(1993){Burrows}, {Singh}, {Nousek}, {Garmire}, \&
  {Good}}]{burrows93}
{Burrows}, {Singh}, {Nousek}, {Garmire}, \& {Good} 1993, \apj, 406, 97

\bibitem[{{Edelstein} {et~al.}(2005{\natexlab{a}})}]{edelstein05a}
{Edelstein}, J. {et~al.} 2005{\natexlab{a}}, \apjl, in press (this volume)

\bibitem[{{Edelstein} {et~al.}(2005{\natexlab{b}})}]{edelstein05b}
---. 2005{\natexlab{b}}, \apjl, in press (this volume)

\bibitem[{{Finkbeiner}(2003)}]{finkbeiner03}
{Finkbeiner}, D.~P. 2003, \apjs, 146, 407

\bibitem[{{Guo} {et~al.}(1995){Guo}, {Burrows}, {Sanders}, {Snowden}, \&
  {Penprase}}]{guo95}
{Guo}, {Burrows}, {Sanders}, {Snowden}, \& {Penprase} 1995, \apj, 453, 256

\bibitem[{{Hartigan} {et~al.}(1987){Hartigan}, {Raymond}, \&
  {Hartmann}}]{hartigan87}
{Hartigan}, P., {Raymond}, J., \& {Hartmann}, L. 1987, \apj, 316, 323

\bibitem[{{Hartmann} \& {Burton}(1997)}]{hartmann97}
{Hartmann}, D. \& {Burton}, W.~B. 1997, {Atlas of galactic neutral hydrogen}
  (Cambridge; New York: Cambridge University Press)

\bibitem[{{Heiles} {et~al.}(1999){Heiles}, {Haffner}, \& {Reynolds}}]{heiles99}
{Heiles}, C., {Haffner}, L.~M., \& {Reynolds}, R.~J. 1999, in ASP Conf. Ser.
  168: New Perspectives on the Interstellar Medium, 211--+

\bibitem[{{Korpela} {et~al.}(2005)}]{korpela05model}
{Korpela}, E.~J. {et~al.} 2005, \apjl, in press (this volume)

\bibitem[{{McGlynn} {et~al.}(1998){McGlynn}, {Scollick}, \& {White}}]{skyview}
{McGlynn}, T., {Scollick}, K., \& {White}, N. 1998, in IAU Symp. 179: New
  Horizons from Multi-Wavelength Sky Surveys, 465--+

\bibitem[{{Murthy} {et~al.}(1993){Murthy}, {Im}, {Henry}, \&
  {Holberg}}]{murthy93}
{Murthy}, J., {Im}, M., {Henry}, R., \& {Holberg}, J. 1993, \apj, 419,
  739

\bibitem[{{Naranan} {et~al.}(1976){Naranan}, {Shulman}, {Friedman}, \&
  {Fritz}}]{naranan76}
{Naranan}, {Shulman}, {Friedman}, \& {Fritz} 1976, \apj, 208,
  718

\bibitem[{{Nishikida} {et~al.}(2005)}]{nishikida05}
{Nishikida}, K. {et~al.} 2005, \apjl, in press (this volume)

\bibitem[{{Paresce} {et~al.}(1983){Paresce}, {Bowyer}, \&
  {Jakobsen}}]{paresce83}
{Paresce}, F., {Bowyer}, S., \& {Jakobsen}, P. 1983, \aap, 124, 300

\bibitem[{{Reynolds} \& {Ogden}(1979)}]{reynolds79}
{Reynolds}, R.~J. \& {Ogden}, P.~M. 1979, \apj, 229, 942

\bibitem[{{Ryu} {et~al.}(2005)}]{ryu05}
{Ryu}, K. {et~al.} 2005, \apjl, in press (this volume)

\bibitem[{{Schlegel} {et~al.}(1998){Schlegel}, {Finkbeiner}, \&
  {Davis}}]{schlegel98}
{Schlegel}, D.~J., {Finkbeiner}, D.~P., \& {Davis}, M. 1998, \apj, 500, 525

\bibitem[{{Seon} {et~al.}(2005)}]{seon05}
{Seon}, K.~I. {et~al.} 2005, \apjl, in press (this volume)

\bibitem[{{Snowden} {et~al.}(1995)}]{rosatASS}
{Snowden}, S.~L., {et~al.} 1995, \apj, 454, 643

\bibitem[{{Welsh} \& {Lallement}(2005)}]{welsh05}
{Welsh}, B.~Y. \& {Lallement}, R. 2005, \aap, 436, 615

\end{thebibliography}

\end{document}